\documentclass{aastex62}

\hypersetup{linkcolor=blue,citecolor=blue,filecolor=blue,urlcolor=black}
\shortauthors{Manek, Brummell \& Lee}

\begin{document}

\title{THE RISE OF A MAGNETIC FLUX TUBE IN A BACKGROUND FIELD: SOLAR HELICITY SELECTION RULES}

\correspondingauthor{Bhishek Manek}
\email{bmanek@ucsc.edu}

\author[0000-0002-2244-5436]{Bhishek Manek}

\author[0000-0003-4350-5183]{Nicholas Brummell}
\author[0000-0002-8229-3040]{Dongwook Lee}
\affil{Department of Applied Mathematics and Statistics, Jack Baskin School of Engineering, University of California Santa Cruz, \\ 1156 High Street, Santa Cruz, California 95064, USA}



\begin{abstract}

The buoyant transport of magnetic fields from the solar interior towards the surface plays an important role in the emergence of active regions, the formation of sunspots and the overall solar dynamo.  Observations suggest that toroidal flux concentrations often referred to as ``flux tubes", rise from their region of initiation likely in the solar tachocline towards the solar surface due to magnetic buoyancy.  Many studies have assumed the existence of such magnetic structures and studied the buoyant rise of an isolated flux tube in a quiescent, field-free environment.  Here, motivated by their formation \citep{2003ApJ..588..630, 2002MNRAS.. L73..L76}, we relax the latter assumption and study the rise of a toroidal flux tube embedded in a large-scale poloidal background magnetic field.  We find that the presence of the large-scale background field severely affects the dynamics of the rising tube.  A relatively weak background field, as low as $6\%$ of the tube strength, can destroy the rise of a tube that would otherwise rise in the absence of the background field.  Surprisingly, the rise of tubes with one sign of the twist is suppressed by a significantly weaker background field than the other.  This demonstrates a potential mechanism for the selection of the preferred helicity of rising and emerging tubes for the solar case that is commensurate with many features of the hemispherical rule.

\end{abstract}



\section{Introduction} \label{sec:intro}

The main observational evidence for a solar dynamo comes from the visible emergence of large-scale magnetic flux at the solar surface as active regions containing sunspots.  The nature of these emergences has dictated much of the dynamics incorporated into theoretical models of dynamo activity.  In particular, the geometry and bi-polar nature of a pair of sunspot has been interpreted as the emergence of a tubular loop of a large-scale strong toroidal magnetic field through the solar surface.

Significant attention has been paid recently (see \cite{Pevtsov_review}) to observations of the magnetic helicity, $ H_{m}=\int_{V} \textbf{A} \cdot (\nabla \times \textbf{A})~dV $ (where $\textbf{A}$ is the magnetic potential, $\nabla \times \textbf{A} = \textbf{B}$), and the current helicity, $ H_{c} = \int_{V} \textbf{B} \cdot (\nabla \times\textbf{B})~dV $ at the solar surface.  These quantities measure the total twist, kinking and linking  in a volume, and give information about the topology of the field that might be important to subsequent processes \citep{Berger_Ruz2000, Rom_Zuc2011}.  The magnetic helicity has dynamic importance, since it is a conserved quantity in ideal magnetohydrodynamics, but is problematical to calculate, whereas the current helicity (or at least some region average of the vertical component) can be readily measured from vector magnetograms of the solar surface.  After many observational studies (see  \cite{Pevtsov_review} for a review), a ``hemispheric helicity rule'' has emerged where the sign of the observed current helicity in many magnetic structures (but, in particular, in active regions) exhibit predominantly negative helicity in the northern hemisphere and positive helicity on the southern hemisphere.  The rule is somewhat weak, with, for example, only 60-75\% of active regions obeying the trend, and is generally independent of the solar cycle, although some recent studies address a brief reversal of the rule in the declining phases of the cycle (e.g. \cite{Tiwari_etal2009,Hao_Zhang2011}).  Since the measured helicity of active regions must be strongly tied to the internal helicity of the emerging toroidal magnetic field, this paper examines that process and posits a possible explanation for these observations.

The creation of the strong large-scale toroidal flux 
is believed to occur in the solar tachocline, the region of strong shear between the convection zone and the radiative zone deep in the solar interior.  
Such shear can take a weak large-scale poloidal field and convert it into a much stronger toroidal field, localized on the scale of the shear.  Simple modeling \citep{JFluidMech..196..323, HWPM1997, 2008ApJ..686..709} has then shown that such configurations can be subject to instabilities that naturally produce arching tube-like toroidal magnetic concentrations that rise against gravity.  The instabilities and rise are driven by magnetic buoyancy \citep{1975ApJ..198..205}, wherein the contribution of the magnetic pressure to the total pressure decreases the gas density within a magnetic concentration and therefore makes it buoyant (assuming that mechanical and thermal equilibrium are likely quickly attained).  

Many studies have been devoted to the vertical transport of magnetic structures ignoring the issue of their creation \citep{Schussler...2d, Moreno...Emonet1996, 1996ApJ...464..999, Fan...2D, Emonet...Insertis...2D}.  In these cases, idealized flux concentrations are used, often referred to as ``flux tubes''.  Typical assumptions have been that the magnetic concentration looks tubular and that it is an isolated magnetic entity rising in a quiescent and field-free region.  
Flux tube models with zero cross-section but with buoyancy, tension, and drag, known as thin flux tubes, have provided insight into rise times and possible field strengths necessary to match observations \citep{Choudhuri...Gilman}.  More realistic flux tubes with finite cross section have shown that their dynamics are more complicated, involving interaction with the generated vortex wake.  A major result of such models has been that the flux tube field must be substantially twisted (with an azimuthal field strength of the same order as the axial field strength) in order for the structure to rise as a coherent entity \citep{Moreno...Emonet1996}.  This confirms that we might expect emerging flux to be helical, but then begs the question of the origin of the observed helicity selection rules.

We here demonstrate that a potential answer comes from relaxing one of the assumptions of the simplified models.  Simulations that do address the origin of the flux tubes \citep{2008ApJ..686..709} show that the magnetic structures are indeed twisted concentrations of flux that are embedded within a space-filling field.  It has been postulated \citep{2003ApJ..588..630} that the dynamics of such embedded concentrations may be distinctly different from those of isolated flux tubes in a field-free environment.  We here test this hypothesis by embedding a previous highly simplified flux tube model within a large-scale magnetic background.  A mechanism that preferentially selects particular combinations of background field orientation and flux tube twist emerges that agrees surprisingly well with the observations.

\section{Theoretical Model} \label{sec:style}

For our model, in a non-convecting fluid layer, we evolve the dynamics of a horizontal cylindrical flux tube (consisting of both axial and azimuthal field to make it helical) embedded in a large-scale horizontal background magnetic field perpendicular to the tube axis.  In the spherical solar sense, the tube should be thought of as a twisted toroidal tube, and the background field should be thought of as poloidal (see Fig.~1a).  The model is therefore very similar to many previous studies (e.g.~\cite{Moreno...Emonet1996}) and virtually identical to that of \cite{1998MNRAS..298..433} (hereafter referred as HFJ), except for the addition of the background field.  The fluid layer is chosen to mimic roughly the top of the tachocline and the lower convection zone in the deep interior of the Sun:  the domain has a density contrast corresponding to the lower 10\% of the convection zone and it is adiabatically stratified as if it were well-mixed, although convection is not present.  The background poloidal field is also chosen to mimic what might be expected if indeed convection were present: we concentrate horizontal field near the bottom of the domain as if it had undergone magnetic pumping by the turbulent convection \citep{Pumping...Tobias_2001}.  We ignore the origins of these fields and study their evolution away from non-equilibrium initial conditions.

We solve the standard equations of compressible resistive magnetohydrodynamics (MHD) with fluid viscosity and thermal conductivity for the velocity $\textbf{u}=(u_x, u_y, u_z)$, magnetic field $\textbf{B}=(B_x, B_y, B_z)$  and thermodynamic quantities using the FLASH code \citep{2014IJHPCA..28, 2000ApJSS..131..273}. Simulations are carried out in a two-dimensional Cartesian domain, $ x \in [-1,1] $ and $ y \in [0,4] $, with a resolution of $200 \times 400$ points where gravity is in the negative $y$-direction and all quantities are independent of $z$.

Magnetic initial conditions consist of a tube and a background horizontal (but vertically-varying) field (see Fig. 1b-d). The magnetic vector potential
$
\textbf{A} = (0,0,A_z)=(0,0,-q r^2 + K)
$
defines a local two-dimensional divergence-free azimuthal tube field within a local radius $r < r_t$ about the tube center $(x,y)=(0,y_c)$, where $q$ represents the twist, $r_{t}$ is the flux tube radius and $K$ is chosen to ensure continuity.  Therefore, in the tube, $B_{x} = -2q(y-y_{c}), B_{y} = 2qx$ relative to a constant axial field $B_z=1$, as in the $ \alpha=0 $ case of HFJ.  In order to represent a large-scale background poloidal field that has been turbulently pumped to the upper tachocline, we add to this
\begin{equation}
\textbf{B}_{back} (y) = \big( B_{s} \exp \Big(\frac{y_{c}-y}{2H_{b}} \Big),0,0\big),
\end{equation}
where $H_{b}$ is the scale-height of the field and $B_{s}$ represents the strength of the background field relative to the initial axial magnetic field which, crucially, can take either sign to represent the alignment in the $x$-direction of the field.   Note that often we quote $B_s$ as a percentage of the axial field for convenience.

The background fluid stratification is defined using a simple polytropic model where the temperature, density, and pressure are given by
\begin{equation}
T = 1 + \theta y'; \hspace{0.1 cm} \rho = (1+\theta y')^{m};
\hspace{0.1 cm} p = (1+\theta y')^{m+1},
\end{equation}
where $y'=4-y$, $\theta$ is the imposed temperature gradient, and $m$ is the polytropic index.  The insertion of the tube magnetic field into the background stratification can lead to an adjustment of the thermodynamics of the tube to accommodate the addition of the magnetic pressure into the total pressure (gas plus magnetic).  It is generally assumed that the total pressure equilibrates quickly.  The accommodation can then be made by changes to any combination of the temperature and density.  We adopt the method of HFJ and insist that the temperature is continuous horizontally at the edge of the tube but varying inside such that density and total pressure are merely a function of height (see Fig. 1e-g).  This setup leads to a plasma $\beta$ of $O(10)$, as in HFJ, which is likely lower than expected solar values but is nonetheless dominated by the gas pressure.

\begin{figure*}
\plotone{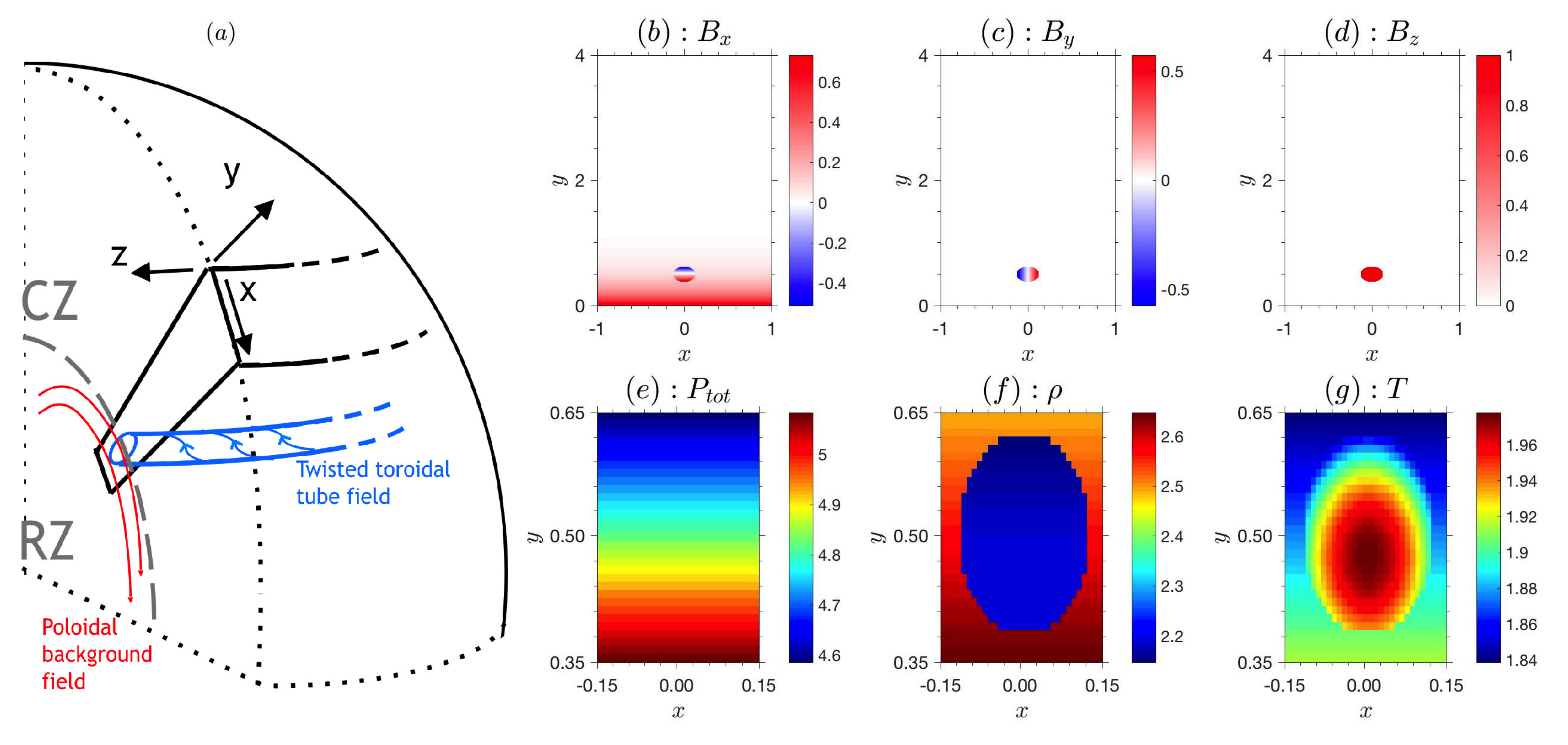}
\caption{Cartoon sketch of the relation of the model to the Sun.  (b-g) shows the initial conditions}


\end{figure*}

We use fixed temperature, stress-free boundaries at the top and bottom, and all other boundary conditions are of outflow type, where the normal derivative is zero.  

We have run a series of simulations of the model described above for a fixed initial configuration except that we vary the strength and orientation of the background magnetic field. The  stratification is weak and adiabatic, specified by $\theta=0.25$ and $m=1.5$.  The (non-dimensional) diffusivities (magnetic, viscous, thermal) governing the equations are $\eta = \mu = 5.e-5$ and $\sigma = 5.e-4$.  The magnetic flux tube is of radius $r_t=0.125$ centered at $(x_c,y_c)=(0,0.5)$ with positive twist given by $q=2.5$ (sufficient to expect a coherent rise).  The background field has a scale height $H_b=r_t=0.125$. The only remaining variables are then the strength of the background field relative to the axial tube strength, $B_s$, and its orientation.   In this paper, we highlight cases in the range $-0.06 \leq B_s \leq +0.20$ (i.e.~$-6\% \leq B_s \leq +20\%$ of the axial field strength).


\begin{figure*}
\plotone{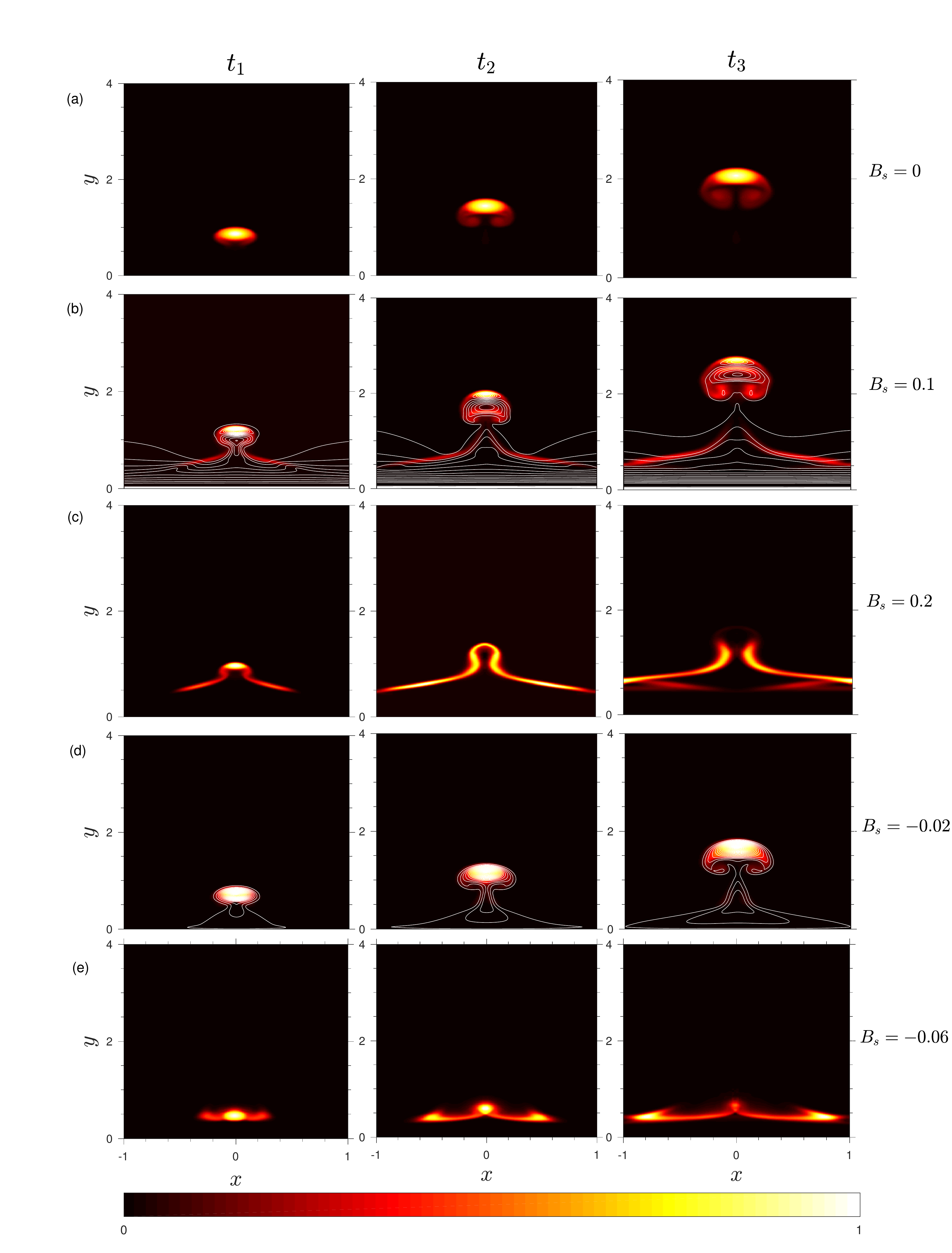}
\caption{Intensity plots of normalised $B_z(x,y,t)$ for $B_{s} = 0,0.1,0.2,-0.02,-0.06$ (a,b,c,d,e respectively) and times $(t_{1}, t_{2}, t_{3})=(5, 10, 15)$ except for (c) where $(t_{1}, t_{2}, t_{3})=(2.5, 5, 8.75)$.  Contours of ${A_z}$ have been added to some panels.  Panels (a,b,d) represent successful rises, whereas in panels (c,e) the rise is suppressed.}
\end{figure*}

\begin{figure*}
\plotone{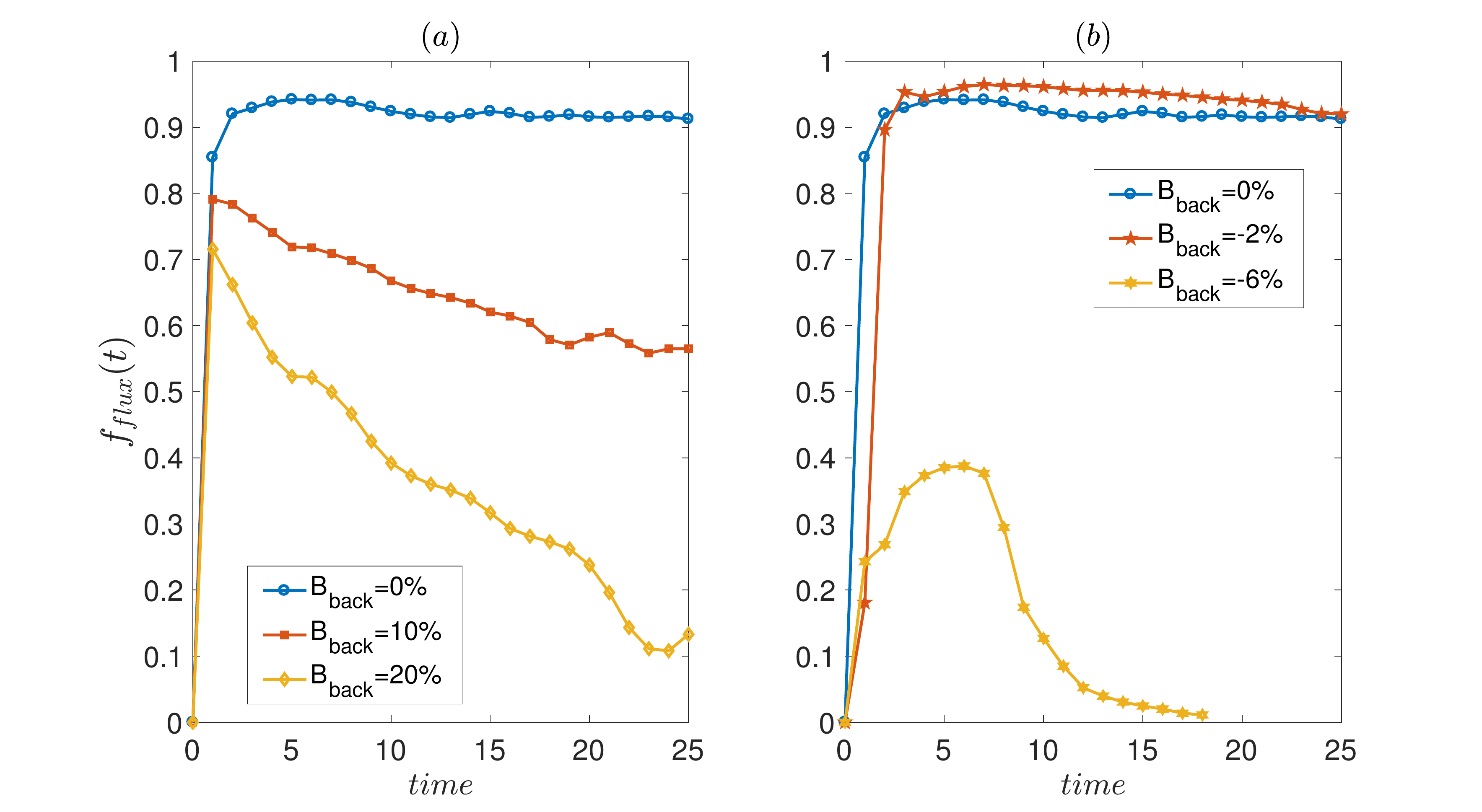}
\caption{Rising flux fraction
$ f_{flux} (t) = {\int\int B_{z,t}^* ~dx dy}/{\int\int B_{z,0} ~dx dy}$ where 
$B_{z,0}=B_{z}(x,y,t=0)$ and $B_{z,t}^*=B_{z} (\{(x,y) ~{\rm where}~ v_{y}(x,y) >v_{y,threshold}\}, t)$ for the same cases as in Fig.~2. 
Here, $ v_{y,threshold} =0.03$ is a judiciously-chosen threshold vertical velocity that filters out unimportant motions.  (a) $B_s \ge 0$.  (b)  $B_s \le 0$.
 }
\end{figure*}

\section{Results} \label{sec:results}
\subsection{Suppression of magnetic structure rise by a background field\label{subsec:suppression}}

Figures 2 and 3 exhibit the main results of this paper.  
Figure 2 shows intensity plots of the evolution of the axial field, $B_{z} $, as a tracer of the magnetic structure, for five different values of $B_s$.  
Figure 3 exhibits $ f_{flux}(t)$, a time trace of the fraction of the initial axial flux that is rising faster than a chosen threshold velocity.  If a simulation exhibits the coherent rise of a magnetic structure, we would expect this measure to be close to unity and relatively constant.

Figure 2(a) shows the results for no background field, $B_s=0$.  
This canonical case exhibits the steady rise of a coherent magnetic structure.  The rise of the tube is initially driven by magnetic buoyancy arising from magnetic perturbations to the density of the initial conditions.  After some time, a pair of counter-rotating trailing vortices develop behind the initial tube, as seen in the tracer $B_z$.  The later rise then becomes dominated by a self-advection of the whole structure driven by the vortices, with only diffusion acting to spread the structure very slowly compared to the rise time.  This result has been well-established by HFJ and others. The coherent rise, in this case, is more quantitively seen as the blue line in Figure 3(a).   After an initial acceleration, most of the initial $B_z$ flux ($\sim 95$\%) rises at a speed above the threshold velocity, indicating that the whole magnetic structure is rising, with only a slight loss of flux (likely due to diffusion).

Figures 2(b) and (c) shows cases where we have included positively-oriented background field with strengths $B_s=0.1, 0.2$ (i.e.,\ 10\% and 20\% of the axial field).   Figure 2(b) exhibits the rise of a still distinct but different tube-like head magnetic structure, but with some axial flux loss to the trailing environment (along ${A_z}$ contour lines).  The head appears to eventually leave the trailing field behind, continuing on to rise seemingly independently as in the no background field case, albeit with a different geometry.  The red line corresponding to this case in Figure 3(a) corroborates these impressions.  The initial magnetically-buoyant acceleration moves a significant (but smaller, $\sim 0.8$) fraction of the flux, but then $f_{flux}$ drops steadily, indicating a regular drainage of axial flux from the rising structure to the non-rising background.  The rate of loss lessens at $t \sim 20$ corresponding to the ``separation'' of the structure from the trailing field (although this behavior is seen more clearly in more diffusive simulations, not shown).  These dynamics may be deemed a successful rise, but just barely.

Figure 2(c), where the background field strength is 20\% of the axial tube field, shows a total suppression of the coherent rise of any structure.  Axial flux in the initial attempted rise is very quickly drained into the trailing environment.  This is confirmed by the yellow line in Figure 3(a) which shows a rapid flux fraction decrease after initiation to low levels ($\sim 0.1$) likely associated solely with waves in the trailing field.  With other simulations, we find that rise is suppressed for a $B_{s}$ somewhere between $16\%$ and $20\%$.  An immediate conclusion is that a relatively weak ($\gtrsim 20$\%) background field will suppress the rise of a twisted tube that would otherwise rise coherently in a non-magnetic background.

Somewhat surprisingly, this conclusion is not independent of the orientation of the background field (relative to the fixed, positive twist (anti-clockwise) of the tube).  Figures 2(d)-(e) and Figure 3(b) show results for $B_s=-0.02, -0.06$ (i.e. 2\% and 6\% of the axial field strength but oriented in the negative direction).   Figure 2(d) and the red line in Figure 3(b) indicate a successful rise with very little flux loss amongst the -2\% background field.  However, Figure 2(e) and the yellow line in Figure 3(b) show a very conclusive failure to rise in the -6\% background field.  It appears that a significantly weaker negatively-oriented field ($|B_s| \gtrsim 6\%$) can suppress the rise of this twisted tube (compared to $|B_s|\gtrsim 20\%$ for a positively-oriented background).

Note that in the simulations presented, we kept the twist of the tube fixed ($q>0$) and switched the orientation of the background field, but this is equivalent to keeping the background field orientation fixed and switching the sign of the twist.  Our results, therefore, suggest a selection rule, where certain intermediate (relative) background field strengths ($6\% \leq |B_s| \leq 20\%$ of the tube strength for the cases actually simulated here) would preferentially allow the emergence of tubes of one sign of twist (where the local azimuthal field at the bottom of the tube is aligned with the background field) over the other.  For example, from Figure 2b, at $B_s = +10\%$ the positively twisted tube simulated rises, whereas at $B_s = -10\%$ the same positively twisted tube is suppressed (since it is already suppressed at $B_s = -6\%$).  This latter configuration is  equivalent to a negatively twisted tube at $B_s = +10\%$, and therefore at $B_s = +10\%$, positively twisted tubes rise whereas negatively twisted tubes do not.


\subsection{Causes of the dynamics\label{subsec:causes}}

\begin{figure*}
\begin{center}
\includegraphics[width=14 cm, height=20 cm]{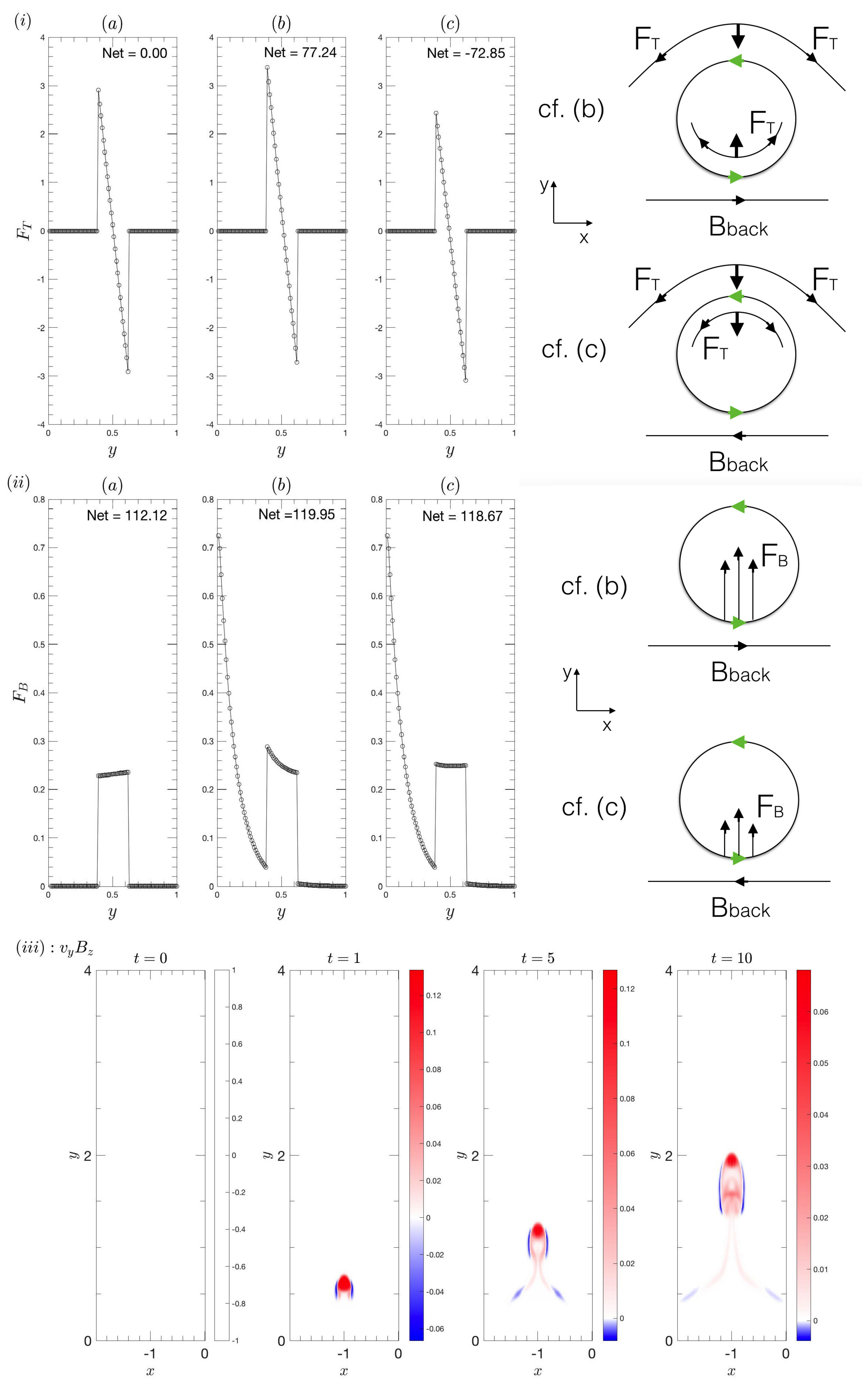}
\end{center}
\caption{Simulation data and cartoons for the forces acting in $y$-direction: (i) tension, $F_{T} = (B_x \partial_x + B_y \partial_y)B_y$ ; (ii) buoyancy force $F_{B} = -\partial_y B_y^2/2 - \rho \theta (m+1) - \partial_y P_{gas}$ evaluated at $x=0, t=0$.  Panels (a,b,c) correspond to $ B_{back}=0\%, 6\%, -6\% $ respectively.  The cartoons show end-on views of rising tubes with black net force arrows for the respective forces (tension or buoyancy) on representative field lines inside and outside the tube for the (b) and (c) cases.  The green arrows show the twist of the tube.  (iii)  Time series of the vertical flux of axial field, $F_A=v_{y}B_{z} $ for $B_s=0.1$.}
\end{figure*}

Numerical evidence and cartoon explanations for the causes of these effects are supplied in Figure 4.   

A tube rising upward through an overlying field induces a downward tension force in the background field due to the wrapping of the field lines around the structure (see the ${A_z}$ contours in Fig.2).  The wrapping also leaves stretched and therefore strong, high tension field in the wake that both resists the formation of the trailing vortices (that would eventually drive the steady coherent rise in the case lacking a background field) and also provides a channel for rapid advective drainage of the axial flux out of the tube thereby reducing its buoyant driving (seen as the negative fluxes in Fig. 4(iii)).
These effects all counter or reduce the upward forces on the structure, and can completely suppress the rise, depending on the relative strengths of the overlying and tube magnetic fields (and the effects of magnetic diffusivity, fixed here). 

Why one orientation of the background field is more efficient at suppressing the rise than the other depends on the contribution of the background field to the internal forces in the tube.  Figure 4 also shows (i) the vertical component of the magnetic tension force $F_{T}$, and, (ii) the vertical component of the total buoyancy force $F_{B}$ on cuts through the initial tube at $x=0$ for relative background fields strengths of 0\%, 6\% and -6\%. The vertical tension force in the 0\% case is symmetrical about the tube center and thus has no net value:  the only role of tension here is to act radially inwards to maintain coherency of the structure. When background field is present, for our chosen (positively) twisted tube, a positively-oriented background field skews the total field inducing stronger positive (upward) magnetic tension at the bottom of the tube and weaker negative (downward) forces at the top, thus inducing a net upward tension 
force in the tube itself.  Reversing the background field (for our fixed tube twist) reverses the asymmetry and induces downward net internal tension.   The cartoons on the right of Figure 4 (i) 
show this pictorially.  The buoyancy force, on the other hand, is enhanced to a similar net value by the presence of either background field direction, although the bottom of the tube is emphasized for $B_s > 0$. The incorporation of a positive background field, therefore, produces net tension and buoyancy forces in the tube that counter the detrimental effect of the magnetic tension induced by the background field during the rise, leading to an enhanced ability to rise.  Conversely, a negative background field creates internal tension forces that act in concert with those induced by the overlying field, to enhance suppression.

\section{DISCUSSION AND CONCLUSIONS\label{subsec:discussion}}

Our simulations have demonstrated that tubes with a twist where the local tube azimuthal field at the bottom of the tube aligns with the background field are more likely to rise than those with the opposite twist since the rise of tubes with the latter alignment is suppressed by a relatively weaker background field.
This selection mechanism is commensurate with the observed solar hemispheric helicity rule(s) in many ways. Figure 5 demonstrates how to translate our model into the solar context, with Figure 5(a) representing one half of the full 22-year solar cycle with the poloidal field of one orientation, and Figure 5(b) the other with the poloidal field reversed. Differential rotation (black arrows) acts on the large-scale poloidal field (red arrows) to create toroidal field (blue arrows) of opposite signs in the two hemispheres.  As has been suggested by simulations of such formation processes \citep{2008ApJ..686..709}, instabilities of this toroidal field create flux tubes with varying current helicities, derived from the correlation of the local azimuthal field (twist: green arrows) created during the process with the axial (toroidal) field (blue arrows).   The right-hand panels of Figure 5 then show which of these magnetic configurations are the least suppressed according to our findings.   For the northern hemisphere, this is always negative helicity tubes, whereas for the southern hemisphere this is always positive, in agreement with the observed solar hemispheric rule.   Note that our selection mechanism originally based on twist has been translated into one on helicity in the solar context owing to the enforced link between the poloidal (background field in the model) and toroidal (axial field in the model) field components induced by the solar differential rotation.

Furthermore, since we find that successful emergence depends on the relative strength of the background field and the tube, and we might expect significant fluctuations in either of these in the solar case, a large scatter may then be anticipated in the helicity observations of the hemispherical rule.

Note also that the cycle invariance exhibited in Figure 5 assumes that the poloidal and toroidal fields switch polarity exactly in phase.  If the switching were to be out of phase at all, there would be a brief period where the selection rule would choose the opposite helicity.  A phase lag is not unexpected (e.g.\ \cite{Char_Dynamo}) and this could explain the observation that the hemispherical rule does not hold so well in the declining phase of the cycle \citep{CCN..2004, Tiwari_etal2009, Hao_Zhang2011, Miesch..phase..lag}.


The selection method revealed by our model therefore fits many crucial elements of the solar helicity observations. Interestingly, it does not directly depend on global rotation and only requires a knowledge of the large-scale dynamo fields, rather than details of the turbulent dynamo processes as in other models \citep{CCN..2004, Miesch..phase..lag}.   However, although these other mechanisms could also contribute to the overall helical content.  Simulations not presented here show that the selection mechanism requires a background field that increases sufficiently fast with depth, however.  For a more uniform overlying field, the rise is strongly suppressed for all helicities.   The observed solar hemispheric rule might then be considered as an indicator of the deeper solar interior field configuration.  A detailed description of these other simulations and the influence of the many parameters of this problem (the stratification, the twist $q$, the magnetic and kinetic Prandtl numbers, etc.) will be forthcoming in a following paper.

\bigskip
We thank David Jones for his initial work and the referee for help clarifying the paper.  All simulations were performed on the Hyades supercomputer at UCSC, acquired under NSF award number AST-1229745. The software used in this work was in part developed by the DOE NNSA-ASC OASCR Flash Center at the University of Chicago.

\begin{figure*}
\epsscale{0.90}
\plotone{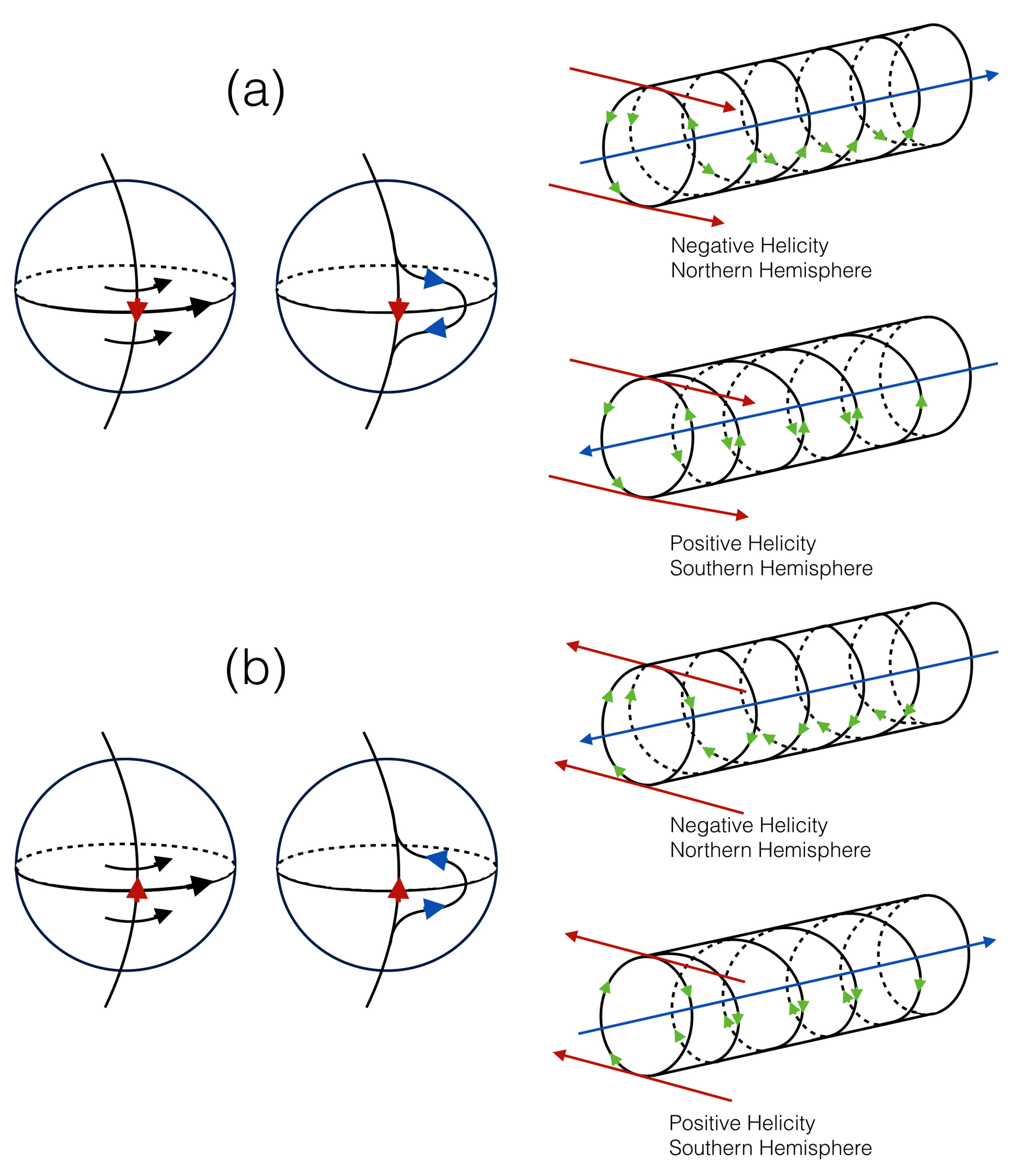}
\caption{Panel (a) and (b) show first and second half of the full 22-year solar cycle respectively. Magnetic field lines are solid lines (overlying poloidal background field as red and toroidal field as blue markers). Green markers indicate the non-axial field, or twist of the tube. The toroidal field is formed by the action of differential rotation (black markers) on the poloidal field. Flux tube configurations that are more likely to rise are shown.}
\end{figure*}




\end{document}